# Bridging Theory with Experiment: Digital Twins and Deep Learning Segmentation of Defects in Monolayer MX$_2$ Phases

Addis S. Fuhr,[1]* Panchapakesan Ganesh,[1] Rama K. Vasudevan,[1] Bobby G. Sumpter,[1]*

[1]Center for Nanophase Materials Sciences, Oak Ridge National Laboratory, Oak Ridge, TN, United States
Correspondence: fuhras@ornl.gov, sumpterbg@ornl.gov

**Abstract**
Developing methods to understand and control defect formation in nanomaterials offers a promising route for materials discovery. Monolayer MX$_2$ phases represent a particularly compelling case for defect engineering of nanomaterials due to the large variability in their physical properties as different defects are introduced into their structure. However, effective identification and quantification of defects remains a challenge even as high-throughput scanning tunneling electron microscopy (STEM) methods improve. This study highlights the benefits of employing first principles calculations to produce digital twins for training deep learning segmentation models for defect identification in monolayer MX$_2$ phases. Around 600 defect structures were obtained using density functional theory calculations, with each monolayer MX$_2$ structure being subjected to multislice simulations for the purpose of generating the digital twins. Several deep learning segmentation architectures were trained on this dataset, and their performances evaluated under a variety of conditions such as recognizing defects in the presence of unidentified impurities, beam damage, grain boundaries, and with reduced image quality from low electron doses. This digital twin approach allows benchmarking different deep learning architectures on a theory dataset, which enables the study of defect classification under a broad array of finely controlled conditions. It thus opens the door to resolving the underpinning physical reasons for model shortcomings, and potentially chart paths forward for automated discovery of materials defect phases in experiments.

**Main Body**
Monolayer MX$_2$ phases (where M is a transition metal and X is either a chalcogen or oxide) have a broad range of potential applications.[1] Transition metal dichalcogenides (TMDs) are the most studied class of these materials, and have optoelectronic properties ranging from metallic to insulating, and can have correlated and topological phases that lead to interesting phenomena such as charge density waves, spin and valleytronics, and superconductivity.[2-8] These materials are highly susceptible to defect formation, which markedly impacts their properties, and can either enhance or degrade device performance.[9-15] The synthesis-structure-property relationships that guide these changes in performance are often abstruse, and difficult to determine. For example, anion vacancies can either increase contact resistance by decreasing carrier mobility in field-effect transistors (MoS$_2$),[15] improve photodetector energy gain by promoting photon upconversion in photodectors (WS$_2$), or induce room-temperature ferromagnetism (VSe$_2$).[10,13]

The ability to radically alter material properties by controlling the type and quantity of defects remains a rich, but difficult to explore area for materials discovery. Identifying defects experimentally can be challenging, which makes it difficult to pinpoint ideal defect targets and the synthesis routes to produce them.[16] Recently, direct atomic-level resolution measurements of defect structure in nanomaterials have become viable with advances in scanning tunneling microscopy (STM) and scanning transmission electron microscopy (STEM).[16-19] STEM measurements in the annular dark-field (ADF) are particularly powerful due to the phase contrast mechanism, which is determined by atom (Z) number.[20-23]

However, the utility of STEM measurements is still somewhat limited. Manual labeling of atoms by a user-in-the-loop is tedious, and lack of automation limits analysis to a small portion of the sample.[24,25] It is therefore difficult to yield a complete picture of the type or concentration of defects with the same level of clarity that, for example, X-Ray diffraction describes crystal structure or mass spectroscopy composition. Machine learning approaches to STEM analysis have consequently emerged as a highly prospective route to achieve high-throughput analysis.[26-32] However, deep learning still requires a user to label images for model training – a process that is not always straightforward with monolayer materials due to beam-matter interactions.[33-43] Image quality (based on signal to noise, or SNR) at low electron doses is poor. High electron doses can reduce noise and make clearer images, but also lead to sample damage and the artificial creation or migration of defects.[26,30,31,44-49] The interplay of these two effects creates uncertainty in connecting synthesis conditions to defect formation. Specifically, it is often unclear if defects are undetected in low dose images is due to noise, or created by the beam in high dose images. This is particularly challenging for materials discovery where the correlation between synthesis method and structure, and resulting structure-property relationships are necessary for designing material-defect systems with tailored functionality.

Here, we address these issues by using first principles calculations to create a digital twin STEM image dataset for 2D monolayer $MX_2$ phases with defects. While high-throughput computational defect-databases have been previously developed with the goal of discovering new materials or identifying trends in properties, we develop computational defect-databases with the goal of improving defect segmentation in microscopy-based experiments for rapid material characterization.[50-54] Several deep learning architectures are trained, and their image segmentation performance evaluated for atomic defects. DFT calculations are used to optimize parent monolayer $MX_2$ phases with various defects, which are then used to generate digital twins through multislice *ab initio* STEM simulations (**Fig. 1**). Our dataset is diverse and comprises of a variety of atomic environments including multiple defect types in the same parent material (e.g., **Fig. 1a-d** shows $NbS_2$ with antisites, transition metal dopants, transition metal adatoms, and vacancies), or the same defect type in multiple materials (e.g., **Fig. 1e-h** shows anion vacancies in $MoSe_2$, $TaSe_2$, $TiTe_2$, and $WSe_2$). In each case, unlike experiment, the precise identity and position of each defect is known *a priori* to labeling images, which allows for thorough benchmarking of various deep learning approaches for STEM. By altering image simulation conditions, we can directly compare the performance of different deep learning architectures on our approximately 600-structure dataset. Thus, showcasing the effectiveness of digital twins for elucidating the basic physics and chemistry limitations for identifying and quantifying defects in monolayer $MX_2$ materials.

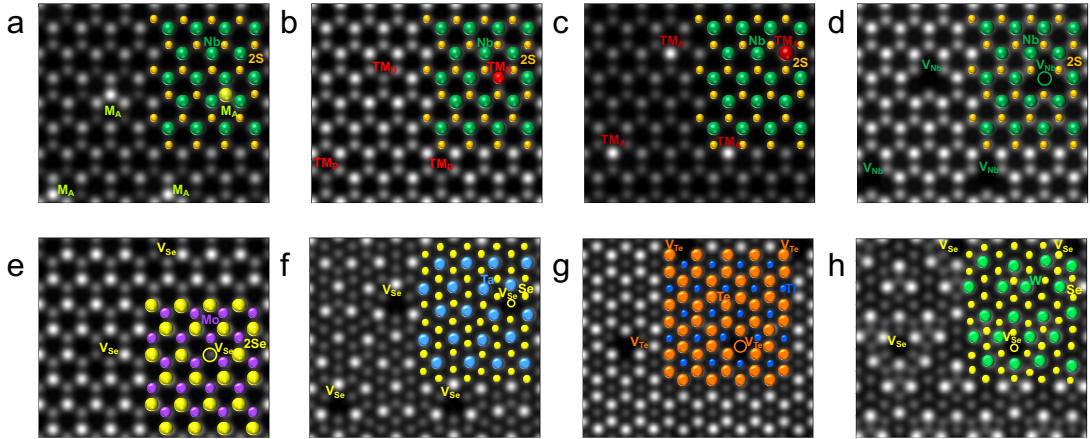

**Fig. 1.** (a-d) STEM digital twins for various defects in NbS$_2$, (e-h) and for anion vacancies in different lattices.

Over 50 monolayer parent structures were obtained from the computational 2D materials database (C2db),[55,56] and converted into 48 atom supercells for defect exploration. We studied 9 different defect classes: anion and metal vacancies ($V_A$ and $V_M$), metal and anion antisites ($M_A$ and $A_M$), metal and anion interstitials ($M_i$ and $A_i$), transition metal dopants and adatoms ($TM_D$ and $TM_A$), and metal antisites with an additional anion vacancy ($M_{2A}$). Chalcogens readily diffuse out of TMD lattices due to their high mobility.[57] Anion vacancies are consequently the most observed defects in TMDs. Hence, for anion vacancies we also varied their concentration (e.g., 1 to 4 vacancies in the supercell), and configurations (e.g., randomly distributed vs. clustered such as in vacancy triplets). Each structure was relaxed using the Perdew–Burke–Ernzerhof, or PBE method with periodic boundary conditions in the Vienna Ab-initio Simulation Package (VASP, see supplementary material for all computational details).[58-62]

Multislice algorithms were used in the *ab inito* TEM (abTEM)[63] package to generate STEM digital twins. DFT periodic boundary conditions lead to somewhat unrealistic images where defect positions perfectly repeat across the larger supercells used in abTEM simulations. We accounted for this by augmenting our dataset, and randomizing defect placement via random cropping, flipping, and rotating of images. While these help us obtain more realistic images, it should be noted that experimental conditions for $MX_2$ phases vary and there are no universally applicable simulation conditions. Our chosen simulation conditions (e.g., probe energy of 80 keV) are not atypical though, and therefore can still provide valuable insight into defect characteristics.

Deep learning models were then used for defect segmentation on our STEM digital twins. As a simple test case, we first construct a generic U-Net (G-UNet) using a similar architecture to standard medical image segmentation tasks (**Fig. 2a**).[64] The encoding pathway consists of a typical stack of convolutional and max pooling layers that capture image context before passing the data through a symmetric expanding (decoding) pathway, which enables precise localization using transposed convolutions. In Fig. **2b-d**, we show heatmaps for G-UNet predictions of different defect types under binary classification (a single defect type vs. the defect-free lattice) conditions. As mentioned earlier, we augmented the dataset to remove the artificial ordering of defects by periodic boundary conditions. This approach is experimentally realistic, and **Fig. 2b-d** shows that our model is not simply memorizing artificially ordered defects from DFT calculations.

Specifically, G-UNet recognizes metal vacancies in different positions and local environments (**Fig. 2b**), anion vacancies with varied concentrations and configurations (e.g., singlets, doublets, and triplets in **Fig. 2c**), and non-vacancy defects in different atomic environments (**Fig. 2d**).

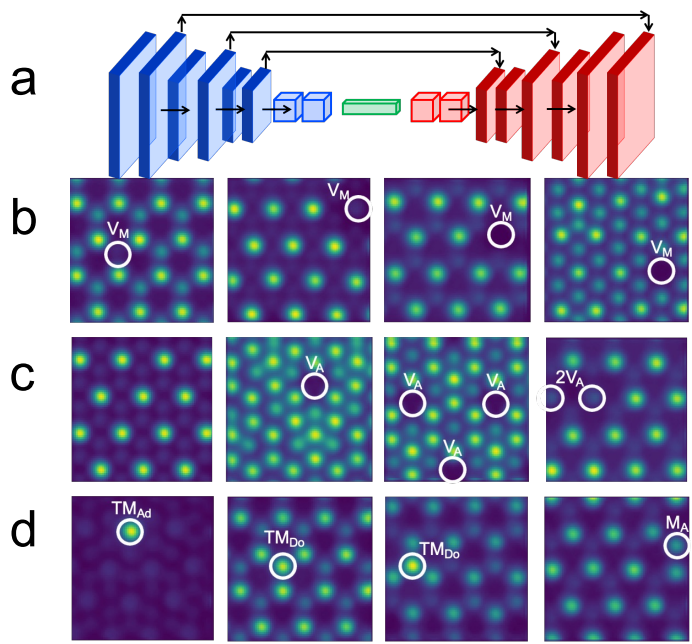

**Fig. 2.** (a) G-UNet architecture. (b-d) Heatmaps for G-UNet predictions on the augmented STEM digital twin dataset.

After observing that G-UNet produced reasonable segmentation heatmaps for our augmented dataset, we conducted a quantitative evaluation of its performance against two other deep learning architectures: an attention UNet (A-UNet) and ResNet-152 (**Fig. 3**). A-UNet modifies G-UNet by incorporating attention modules in the skip connections between encoding and decoding pathways to selectively ignore irrelevant features and focus on salient features.[65] The resulting attention maps have been shown to improve segmentation performance by designating important pixels, and weighing their features in the skip connections. ResNet-152 is an advanced segmentation architecture designed by Microsoft[66] that utilizes convolutional neural networks with skip connections to create residual networks (ResNets). The skip connections connect the input to multiple output layers and enables gradient information to bypass some of the layers and flow directly into earlier layers. This process preserves information from earlier layers, prevents information loss in later layers, and yields deeper networks that learn more complex patterns in images to improve segmentation.

In **Fig. 3a,b** we compare each of these models under binary classifications conditions based on their intersection over union (IOU, **Fig. 3a**) and their F1-scores (**Fig. 3b**). Our dataset is imbalanced, as anticipated in experiments where certain defects are expected to occur more frequently (e.g., anion vacancies) than others (e.g., metal interstitials). We therefore examine the relationship between each score and total number of defect instances and find no correlation between defect count and segmentation performance. This result indicates that variations in architecture performance are due to physical differences in defect type and their corresponding representation in a STEM image instead of availability of training data.

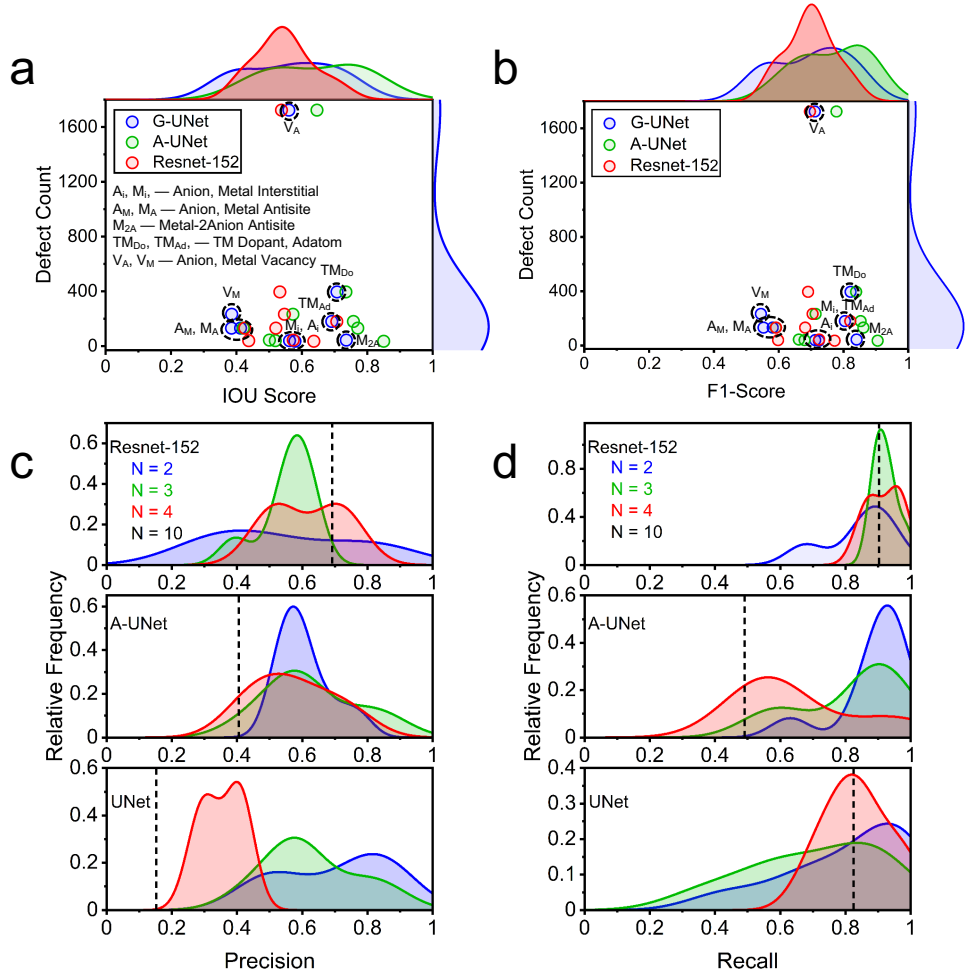

**Fig. 3.** (a,b) Segmentation performance for 3 deep learning architectures based on IOU score (a) and F1-score (b) under binary classification conditions. Line plots represent data distribution. (c,d) Performance based on precision (c) and recall (d) as the number of classes increase. All classification tasks include the defect-free lattice (e.g., N=2 indicates binary classification of one defect class vs. the defect-free lattice). For N=10, all of the defect classes are included and a dashed black line is used instead of a distribution.

It is not uncommon for experimental segmentation tasks to encounter unidentified features such as unexpected impurities. In the context of our digital twin dataset, binary classification tasks mimic these conditions by requiring the deep learning model to recognize and distinguish each defect from the defect-free-lattice while encountering other impurities. Moreover, we have also included several additional defect classes in our dataset that are not labelled. Many of these visually resemble other defects such as Li-dopants, which are undetectable in STEM and resemble metal vacancies (see supplementary material). A-UNet (F1$_{Avg}$ = 0.76) outperforms G-UNet (F1$_{Avg}$ = 0.68) and ResNet-152 (F1$_{Avg}$ = 0.70) in binary classification, which otherwise have similar performance (**Fig. 3a,b**). It is not necessarily surprising that an advanced architecture such as ResNet-152 did not significantly outperform G-UNet. The large number of parameters for ResNet-152, and its focus on learning intricate features for a variety of classes does not always significantly improve segmentation under binary conditions. A-UNet, on the other hand, has the best performance, which is likely due to hyperfocus on pertinent features for each defect class, and suppressing features applicable to the other defect types.

In **Fig. 3**, we also explored the performance of each model as the number of defect classes (N) increase, and the complexity of the segmentation task grows. A-UNet maintains a base higher-level performance than G-UNet, but further increases in N lead to a drop in F1-score for both models (see supplementary materials). Although both models experience a reduction in performance, the underlying causes vary. Much of the performance reduction for G-UNet stems from lost precision (**Fig. 3c**) whereas for A-UNet it is lost recall (**Fig. 3d**). As described earlier, attention mechanisms weigh the importance of each pixel, and improve segmentation by learning more intricate features for target classes, while ignoring irrelevant ones.[65] In this case, the relatively simple G-UNet model achieves stable or modestly improved recall with increasing N (false negatives), but struggles with learning complex features resulting in a dramatic drop in precision (false positives). Introducing attention addresses the precision issue by prioritizing intricate features, but at the cost of lower recall.

While A-UNet had the best performance under binary classification conditions, the deeper architecture of ResNet-152 proves valuable as complexity grows. The average F1-score for ResNet-152 is almost unchanged when N increases from 2 to 3 (see supplementary materials), which leads to ResNet-152 and A-UNet having similar average F1-Scores (0.70 vs. 0.68, respectively). As N increases further, ResNet-152 is the only model that improves segmentation performance, and it begins to outperform A-UNet when N is greater than 4. Thus, as shown with other segmentation tasks,[66] the deep architecture of ResNet-152 leads to a greater ability to capture a larger number of general features of the input images, learn more subtle differences between classes, and improve performance across all metrics when dealing with large, complex, multiclass segmentation tasks.

Atomic variability, and the number of labelled and unlabeled defect classes are not the only challenges for defect classification in experiments. Sample rotations and grain boundaries can also complicate segmentation tasks.[67] Moreover, as described earlier, image quality is strongly related to electron dose, and high doses can lead to beam damage-induced defect formation.[26,30,31,44-49] We therefore expand our dataset to include "messy" STEM images (**Fig. 4**). This messy dataset includes the base images described in previous sections (e.g., with varied defect types, environments, and concentrations as shown in **Fig. 4a-c** with $V_{Te}$ in $VTe_2$), but also introduces random rotations (e.g., as shown with $Mo_{0.94}V_{0.06}S_2$ rotated by 45 degrees in **Fig. 4d**), grain boundaries (or, GB as shown with $Mo_{0.94}S_2$ in **Fig. 4e**), and beam damage (or, BD as shown in **Fig. 4f**). Rotations, GBs, and BD are quasi-randomly generated (e.g., GBs can be randomly separated or overlap within a 2-3 Å range, see supplementary materials). Messy dataset images are reproduced at multiple electron doses, which are simulated by Poisson noise (e.g., at 1e4 $e/Å^2$ to 5e7 $e/Å^2$ in **Fig. 4g-i**).

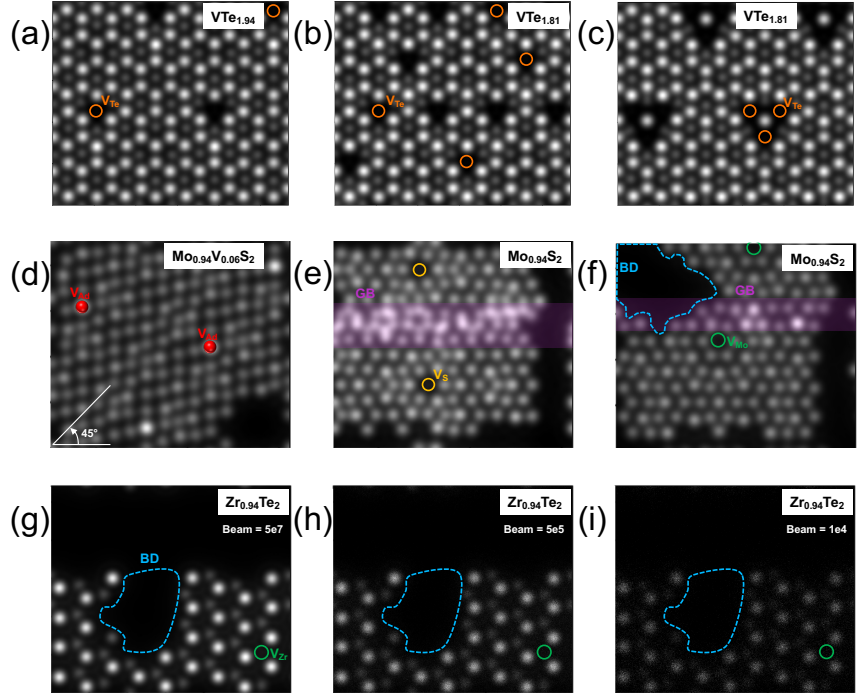

**Fig. 4.** (a-c) An example of the base complexity of the non-messy dataset: anion vacancies at low concentration (a), higher concentrations with disorder (b), and higher concentrations with order (c). (d-i) Messy dataset example images where (d) is a 45 degree rotated structure, and (e,f) are structures with grain boundaries (e) and beam damage (f). (g-i) Simulated effects of electron dose on image quality.

We evaluate segmentation performance on the messy dataset by plotting the IOU (**Fig. 5a**) and F1-scores (**Fig. 5b**) as a function of electron dose. The performance for ResNet-152 on the high electron dose messy dataset is nearly identical to the clean, no-noise dataset. However, performance drops significantly for all architectures as the electron dose decreases, and image quality deteriorates. While the exact performance varies for each architecture, the scatter plot for the dose-dependent reduction in IOU and F1-score can be fitted to the same monoexponential decay equation: (eqn. 1: $D = D_0 + Ae(-B/\alpha)$, where $D_0$ is initial detectability without Poisson noise, A is the amplitude of reduction in detectability, B is the beam electron dose, and $\alpha$ the characteristic decay dose at which detectability decays by 1/e).

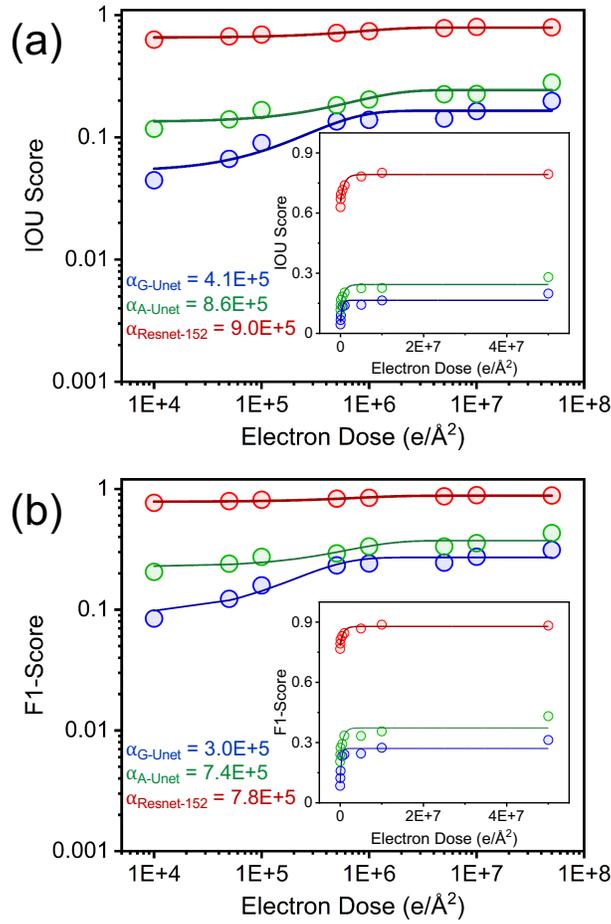

**Fig. 5.** Segmentation performance for the messy dataset under N = 10 conditions. In each case the IOU score (a) and F1-scores (b) are shown as a scatter plot, and the fit to eq. 1 as a solid line. The figure insets represents the data on a linear scale.

STEM experiments use a small probe of electrons to scan across the sample. The atomic number (Z) of each element determines the scattering angle where transmitted electrons are collected by a detector to produce a signal. However, penetration of the sample by the electron beam produces only a limited number of scattering events. The random nature of electron-atom interactions results in statistical fluctuations in the number of detected electrons, which can be modelled as Poisson noise.[63] The average number of signal events per unit area of the sample decreases as electron dose is reduced. As the probability of observing scattering events decreases, the relative contribution of Poisson noise to image quality grows, and atom detectability drops as SNR worsens. If SNR is the ratio between the average signal and the standard deviation in noise, and we are modeling the probability of detecting an electron scattering event as being proportional to a Poisson probability density function, the SNR standard deviation should be the square root of the expected mean number of dose-dependent electron scattering events per atom.

If thermal or other sources of noise are not present, the impact of electron dose on segmentation performance can be assessed by comparing $\alpha$ factors (eqn. 1). This fit parameter is reflective of the threshold dosage required for classifying atom types based on the number of electron scattering occurrences per atom. Low electron doses reduce the likelihood of detecting scattering events, resulting in high SNR and correspondingly low image quality. Hence, the segmentation performance of each model is expected to exponentially decrease with electron dose following the relationship described in eqn. 1. In **Fig. 5**, the dose-dependent IOU and F1-scores obey these relationships for each deep learning model, and their performance are fit with eqn.1 using distinct $\alpha$ terms. These results indicate that our approach can distinguish the performance of each model based on the actual physical relationships that affect image quality. Specifically, the performance degradation of each architecture at lower electron doses is directly related to the lower probability of detecting scattering events, which was modeled as Poisson noise. In addition, the performance degradation for each model based on the $\alpha$ factor is distinct, and reflective of their shortcomings for classifying defects based on the reduced probability of detecting scattering events. Here, we find that A-UNet and ResNet-152 have distinct base detectability but exhibit similar noise performance whereas G-UNet has much poorer performance.

In conclusion, we have demonstrated that deep learning segmentation models can be systematically evaluated for defect classification using STEM digital twins. Attention mechanisms were found to be most effective in scenarios where classification tasks involved fewer target defect classes to identify but complicated by the presence of numerous unclassified impurities. ResNet-152, conversely, had the best performance when dealing with a plethora defect classes in complex multiclass segmentation tasks. More importantly, we observed no correlation between performance and defect instance count in an unbalanced dataset with unevenly represented defect classes, and successfully returned the Poisson noise function for each architecture when evaluating dose-dependent performance. Taken together, these results indicate that variations in model performance can be attributed to the expected physical mechanisms that impact image quality in STEM experiments. Materials discovery is not limited to new crystal structures, but also includes understanding how synthesis conditions impact defect formation, and how defects alter structure-property relationships. In both cases, STEM digital twins should play a significant role in improving our ability to classify and quantify defects in monolayer $MX_2$ materials and opens the door for designing materials with tailor-made functionalities for different applications.

**Acknowledgments**
ASF acknowledges support from the Alvin M. Weinberg Fellowship at Oak Ridge National Laboratory. This work was carried out at Oak Ridge National Laboratory's Center for Nanophase Materials Sciences, a US Department of Energy Office of Science User Facility, and used the Compute and Data Environment for Science (CADES) at Oak Ridge National Laboratory, which is supported by the Office of Science of the DOE under Contract DE-AC05-00OR22725.

**Data Availability**
The data that support the findings of this study are openly available in [Defects in monolayer $MX_2$ dataset], reference number [TBA].

23. D. Rhodes, D. A. Chenet, B. E. Janicek, C. Nyby, Y. Lin, W. Jin, D. Edelberg, E. Mannebach, N. Finney, A. Antony, T. Schiros, T. Klarr, A. Mazzoni, M. Chin, Y. c. Chiu, W. Zheng, Q. R. Zhang, F. Ernst, J. I. Dadap, X. Tong, J. Ma, R. Lou, S. Wang, T. Qian, H. Ding, R. M. Osgood, Jr., D. W. Paley, A. M. Lindenberg, P. Y. Huang, A. N. Pasupathy, M. Dubey, J. Hone, and L. Balicas, Nano Letters **17** (3), 1616 (2017).
24. W. Li, K. G. Field, and D. Morgan, npj Computational Materials **4** (1), 36 (2018).
25. S. I. Molina, D. L. Sales, P. L. Galindo, D. Fuster, Y. González, B. Alén, L. González, M. Varela, and S. J. Pennycook, Ultramicroscopy **109** (2), 172 (2009).
26. S.-H. Yang, W. Choi, B. W. Cho, F. O.-T. Agyapong-Fordjour, S. Park, S. J. Yun, H.-J. Kim, Y.-K. Han, Y. H. Lee, K. K. Kim, and Y.-M. Kim, Advanced Science **8** (16), 2101099 (2021).
27. K. Lee, J. Park, S. Choi, Y. Lee, S. Lee, J. Jung, J.-Y. Lee, F. Ullah, Z. Tahir, Y. S. Kim, G.-H. Lee, and K. Kim, Nano Letters **22** (12), 4677 (2022).
28. M. Ziatdinov, O. Dyck, X. Li, B. G. Sumpter, S. Jesse, R. K. Vasudevan, and S. V. Kalinin, Science Advances **5** (9), eaaw8989 (2019).
29. C.-H. Lee, A. Khan, D. Luo, T. P. Santos, C. Shi, B. E. Janicek, S. Kang, W. Zhu, N. A. Sobh, A. Schleife, B. K. Clark, and P. Y. Huang, Nano Letters **20** (5), 3369 (2020).
30. K. M. Roccapriore, M. G. Boebinger, O. Dyck, A. Ghosh, R. R. Unocic, S. V. Kalinin, and M. Ziatdinov, ACS Nano **16** (10), 17116 (2022).
31. A. Maksov, O. Dyck, K. Wang, K. Xiao, D. B. Geohegan, B. G. Sumpter, R. K. Vasudevan, S. Jesse, S. V. Kalinin, and M. Ziatdinov, npj Computational Materials **5** (1), 12 (2019).
32. Y. Guo, S. V. Kalinin, H. Cai, K. Xiao, S. Krylyuk, A. V. Davydov, Q. Guo, and A. R. Lupini, npj Computational Materials **7** (1), 180 (2021).
33. S. Ning, T. Fujita, A. Nie, Z. Wang, X. Xu, J. Chen, M. Chen, S. Yao, and T.-Y. Zhang, Ultramicroscopy **184**, 274 (2018).
34. J.-H. Shim, H. Kang, S. Lee, and Y.-M. Kim, Journal of Materials Chemistry A **9** (4), 2429 (2021).
35. J.-H. Shim, Y.-H. Kim, H.-S. Yoon, H.-A. Kim, J.-S. Kim, J. Kim, N.-H. Cho, Y.-M. Kim, and S. Lee, ACS Applied Materials & Interfaces **11** (4), 4017 (2019).
36. S. Van Aert, A. J. den Dekker, D. Van Dyck, and A. van den Bos, Ultramicroscopy **90** (4), 273 (2002).
37. H.-P. Komsa, S. Kurasch, O. Lehtinen, U. Kaiser, and A. V. Krasheninnikov, Physical Review B **88** (3), 035301 (2013).
38. G. Algara-Siller, S. Kurasch, M. Sedighi, O. Lehtinen, and U. Kaiser, Appl. Phys. Lett. **103** (20) (2013).
39. K. Elibol, T. Susi, G. Argentero, M. Reza Ahmadpour Monazam, T. J. Pennycook, J. C. Meyer, and J. Kotakoski, Chemistry of Materials **30** (4), 1230 (2018).
40. A. Ghosh, M. Ziatdinov, O. Dyck, B. G. Sumpter, and S. V. Kalinin, npj Computational Materials **8** (1), 74 (2022).
41. D. B. Lingerfelt, P. Ganesh, J. Jakowski, and B. G. Sumpter, Advanced Functional Materials **29** (52), 1901901 (2019).
42. D. B. Lingerfelt, T. Yu, A. Yoshimura, P. Ganesh, J. Jakowski, and B. G. Sumpter, Nano Letters **21** (1), 236 (2021).
43. D. B. Lingerfelt, P. Ganesh, J. Jakowski, and B. G. Sumpter, Journal of Chemical Theory and Computation **16** (2), 1200 (2020).
12

# Supplementary Materials

**Details of Computational Approach**

Plane-wave PAW calculations used the Perdew-Burke-Ernzerhof (PBE) functional in VASP. Supercells for "parent" monolayer $MX_2$ structures were obtained by a 4X4X1 (48 atom) expansion of ~50 monolayer materials selected from the C2db database, which have 15 Å vacuum spacing. Each supercell was relaxed with and without the defects described in the main article. The energy cut-off was 800 eV, and the structures were relaxed until the forces were smaller than 0.02 eV per Å per atom. A Monkhorst-Pack K-point grid of 4X4X1 was used for all optimizations. Multislice electron scattering simulations used the abTEM python package. For all STEM simulations the DFT supercell was orthogonalized leading to an expansion of 4 repeating supercells (192 atoms). All STEM simulations used a probe energy of 80 keV, a semi-angle cutoff of 25 degrees, a defocus value of 40 Å, and a spherical aberration (Cs) value of 4e5 Å. The flexible annular detector was used in all simulations, which allows us to select the integration limits (here, between 50 and 200 for HAADF) for the multislice simulations.

The "messy" dataset used a similar procedure, except rotations, grain boundaries, and/or beam damage conditions were introduced during the orthogonalization step. For each structure, a random choice was made between a rotation with no grain boundaries, or a rotation with grain boundaries. If rotation without grain boundaries was selected, the lattice rotation was randomly selected from 0, 30, or 45 degrees. On the other hand, if grain boundary was chosen, a random *relative* rotation was selected from 0, 10, 30, or 45 degrees. For grain boundaries, in addition to the relative rotation between grains, a grain separation distance was randomly chosen between -3 to 3 Å. Here, negative values indicate one monolayer structure is partially stacked on top of the other (e.g., -3 Å would indicate there is 3 Å of overlap between both grains) and positive values indicate a separation between layers (e.g., 3 Å would indicate one monolayer is 3 Å apart from the other). Once the random choice between grain boundary or grain boundary-free lattice rotation was made, and rotation or grain boundary operation was complete, a second random choice was introduced on whether there would be beam damage or no beam damage (e.g., beam damage could happen in either the rotated cells, or the grain boundary cells). If beam damage was selected, a random atom and several of its nearest neighbors (between 5 and 30 atoms in total) were selected for removal from the structure. Electron dose was simulated for each image in the entire messy dataset (1E4 to 7.5E7 e/$Å^2$) using the Poisson noise function in abTEM.

All machine learning trainings used the Pytorch framework.[68] Prior to training ML, we augmented the dataset to circumvent issues associated with artificial ordering of defects by periodic boundary conditions. Every image was randomly cropped to a size of ~8% of the original supercell. After random cropping, an additional augmentation was selected randomly, and the image was either flipped horizontally, flipped vertically, or rotated by 90 degrees.

The G-UNet model consists of a down-sampling and up-sampling path. The down-sampling path has four consecutive double convolution layers (two 2D convolutional layers with a kernel size of 3 and padding of 1, followed by batch normalization and rectified linear unit, or ReLU activation).

The corresponding feature maps are 64, 128, 256, and 512. A bottleneck layer then uses a double convolution layer to increase the feature map to 1024, before passing through 4 up-sampling blocks. The up-sampling blocks are composed of either convolution transpose layers (with kernel size of 2 and stride of 2) or bilinear up-sampling layers (scale factor of 2) resulting in 512, 256, 128, and 64 sized feature maps. The final layer is a 2D convolutional layer with a kernel size of 1, which performs a pixel-wise classification of the input image.

A-UNet simply modifies G-UNet to include attention by adding an attention block after each down-sampling block. The input for the attention block is the output of the previous down-sampling block and the output of the corresponding up-sampling block. The attention block has three convolutional layers (the first two with a kernel size of 1, and the last with a kernel size of 3), and is followed by batch normalization and ReLu activation. It generates a spatial attention map by weighing features found in the previous down-sampling block before concatenation with the corresponding up-sampling block. ResNet-152 was used with pre-loaded weights from imagenet.[66]

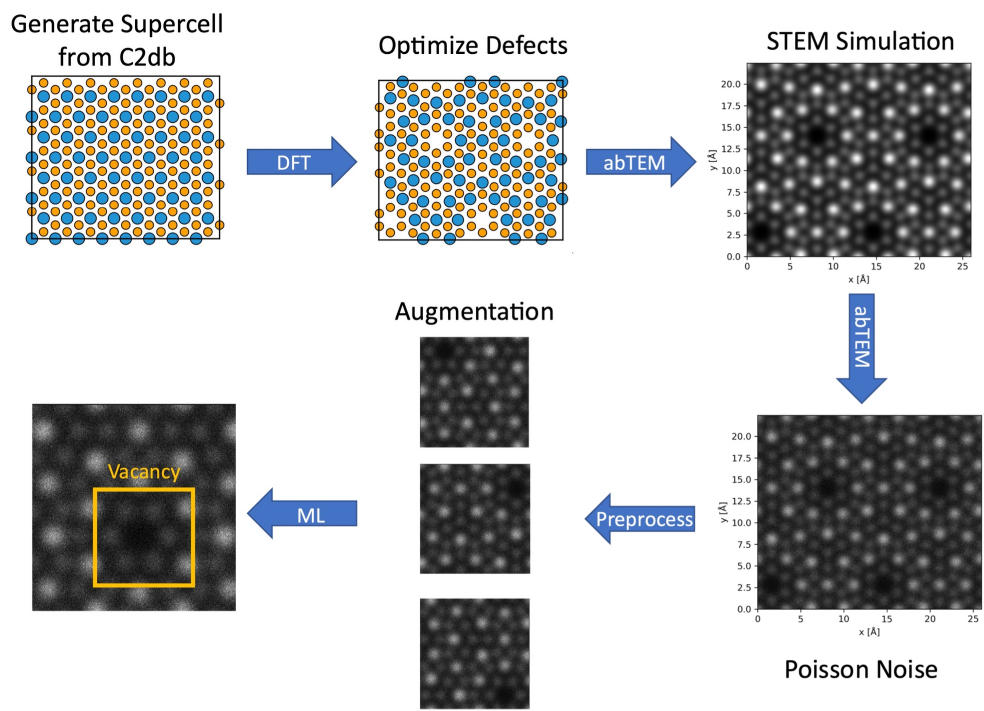

**Fig. S1** Computational workflow for creating STEM digital twins. Step 1) A parent structure is selected from C2db for supercell generation. Step 2) The structure is then optimized with defects using DFT. Step 3) The DFT optimized structure is used in an ab initio electron scattering simulation to generate the digital twin. If making "messy" STEM images, additional operations (rotations, grain boundaries, beam damage) can be done at this step. Step 4) In the messy dataset noise is added to the image to simulate low electron doses. Here, for clarity, we show noise effects on the same non-messy STEM image exemplified in step 3. Step 5) Augmentation (random cropping, flipping, etc.) is used to remove artificial ordering of defects by periodic boundary conditions. Step 6) ML STEM segmentation model is trained.

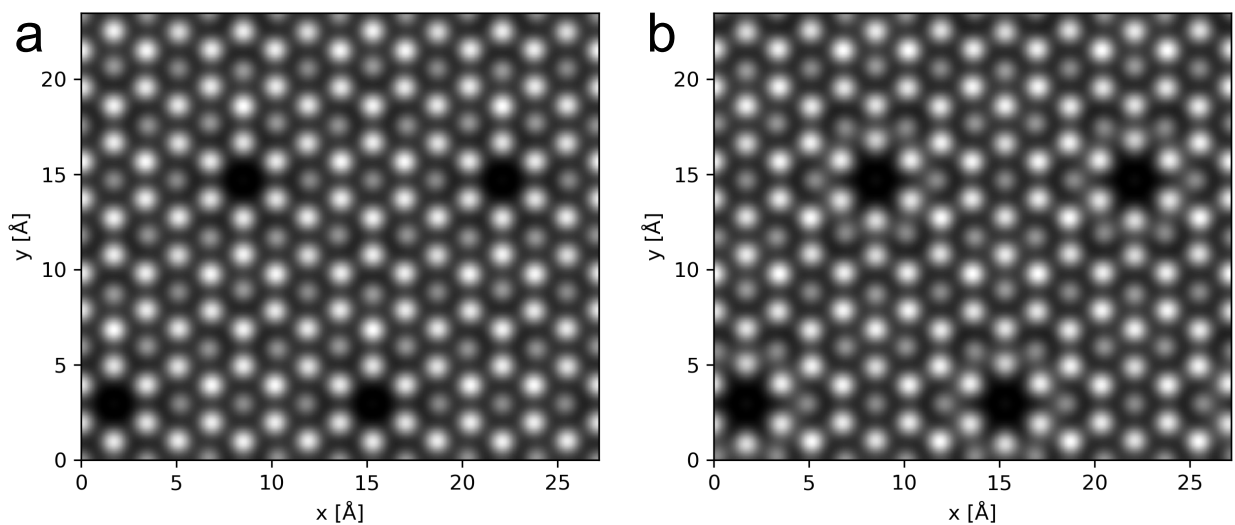

**Fig. S2** Comparing labelled defects versus unlabeled impurities. a) A metal vacancy in $CrSe_2$ is an example of a labelled defect. b) The same $CrSe_2$ structure with an unlabeled impurity (Li dopant). Lithium does not appear in STEM experiments, Li dopants therefore look similar to metal vacancies.

| Model | F1-Score$_{Avg}$ (N=2) | F1-Score$_{Avg}$ (N=3) | F1-Score$_{Avg}$ (N=4) | F1-Score$_{Avg}$ (N=10) |
|---|---|---|---|---|
| G-UNet | 0.68 | 0.57 | 0.49 | 0.32 |
| A-UNet | 0.76 | 0.70 | 0.59 | 0.43 |
| ResNet-152 | 0.70 | 0.68 | 0.73 | 0.88 |

**Table. S1** Average F1-score for the different segmentation as a function of the number of defect classes.